\documentclass[runningheads]{llncs}
\usepackage{graphicx}
\usepackage{framed}
\usepackage[backgroundcolor=blue!10,bordercolor=gray]{todonotes}

\begin{document}

\title{A Comparison of Different Source Code Representation Methods for Vulnerability Prediction in Python}

\titlerunning{Comparison of Source Code Representations}
%
\author{Amirreza Bagheri\inst{1} \and
P\'eter Heged\H{u}s\inst{2,3}
}

\authorrunning{Bagheri and Heged\H{u}s}
%

\institute{
University of Szeged, Software Engineering Department\\
\email{bagheri@inf.u-szeged.hu}
\and
MTA-SZTE Reserach Group on Artificial Intelligence, ELKH\\
\and
FrontEndART Ltd., Szeged, Hungary\\
\email{hpeter@inf.u-szeged.hu}
\vspace{-6pt}
}

\maketitle              

\begin{abstract}
In the age of big data and machine learning, at a time when the techniques and methods of software development are evolving rapidly, a problem has arisen: programmers can no longer detect all the security flaws and vulnerabilities in their code manually.
To overcome this problem, developers can now rely on automatic techniques, like machine learning based prediction models, to detect such issues.
An inherent property of such approaches is that they work with numeric vectors (i.e., feature vectors) as inputs.
Therefore, one needs to transform the source code into such feature vectors, often referred to as \textit{code embedding}.
A popular approach for code embedding is to adapt natural language processing techniques, like text representation, to automatically derive the necessary features from the source code.
However, the suitability and comparison of different text representation techniques for solving Software Engineering (SE) problems is rarely studied systematically.
In this paper, we present a comparative study on three popular text representation methods, word2vec, fastText, and BERT applied to the SE task of detecting vulnerabilities in Python code.
Using a data mining approach, we collected a large volume of Python source code in both vulnerable and fixed forms that we embedded with word2vec, fastText, and BERT to vectors and used a Long Short-Term Memory network to train on them.
Using the same LSTM architecture, we could compare the efficiency of the different embeddings in deriving meaningful feature vectors.
Our findings show that all the text representation methods are suitable for code representation in this particular task, but the BERT model is the most promising as it is the least time consuming and the LSTM model based on it achieved the best overall accuracy(93.8\%) in predicting Python source code vulnerabilities.

\end{abstract}
\section{Introduction}\label{sec:intro}

Security bugs (i.e., vulnerabilities) in software are becoming more and more difficult to identify in today's applications, allowing hackers and attackers to profit from their exploit. Every year, tens of thousands of such vulnerabilities are discovered and fixed.
Manually auditing source code and finding vulnerabilities is costly at best, if not impossible at all. Therefore, researchers and practitioners have proposed various tools that can help in discovering vulnerabilities automatically.
Classical vulnerability detection tools rely on static~\cite{cousot2005astree,arroyo2016user,olesen2014coccinelle} or dynamic~\cite{srivastava2004atom,skaletsky2010dynamic,waddington2010dynamic} code analysis, symbolic execution or taint analysis.
However, with the advent of efficient machine learning techniques, new approaches appear that try to solve Software Engineering (SE) problems by training AI prediction models on large amount of annotated code samples. Vulnerability detection is one typical such SE task that has been addressed with these new ML approaches. As of today, using prediction models to decide if a source code fragment is vulnerable or has became a very common and effective approach. An inherent property of such approaches is that they can work with numeric vectors (i.e., feature vectors) as inputs.Therefore, one needs to transform the source code into such feature vectors, often referred to as \textit{code embedding}.
This process can be either manual (i.e., defining and extracting features from source code manually, like lines of code, number of branches, code complexity) or automatic (i.e., applying ML based techniques to automatically learn the vector representation of code).
A popular approach for automatic code embedding is to adapt natural language processing techniques, like text representation~\cite{wen2016learning}, to automatically derive the necessary features from the source code.Despite their popularity, the suitability and comparison of different text representation techniques for solving SE problems has been rarely studied systematically.Given the fact that the accuracy of prediction models relies heavily on the appropriate representation of input data, we need empirical data about the effect of such representations on the underlying SE task to be solved.

In this paper, we present a comparative study of three popular text representation methods, word2vec~\cite{church2017word2vec}, fastText~\cite{joulin2016fasttext}, and BERT~\cite{devlin2018bert} applied to the SE task of detecting vulnerabilities in Python code.
We applied a data mining approach to collect a suitable training data for training vulnerability prediction models.Using a heuristic approach (i.e., searching for simple terms indicating security fixes in commit logs) we collected a large volume of Python source code from GitHub in both vulnerable and fixed forms.We generated the vector representation of these code fragments using automatic code embedding based on the word2vec, fastText, and BERT text representation methods and used a Long Short-Term Memory~\cite{sundermeyer2012lstm} network to create a vulnerability prediction model based on them. Training the LSTM model with the same architecture on each of the different code representations, we could compare the efficiency of the various embeddings in deriving meaningful feature vectors for vulnerability prediction. We investigated the following two research questions using the above described methodology:

\textbf{RQ1:} Is there a significant difference in the performance of the vulnerability prediction models based on the different code embedding methods?

\textbf{RQ2:} Are some of the code embedding methods more suitable for predicting certain types of vulnerabilities than others?

Our findings show that all the text representation methods are suitable for code representation in this particular task, but the BERT model is the most promising, as it is the least time consuming, and the LSTM model based on it achieved the best overall accuracy(93.8\%) in predicting Python source code vulnerabilities.
Regarding the various vulnerability types, we observed slight variances in model performances based on the applied source code embeddings.
Nonetheless, the prediction model based on the word2vec representation of code clearly outperformed models based on fastText and BERT for detecting SQL injection, while in case of Command Injection, Cross-Site Request Forgery (XSRF), Remote Code Execution (RCE), and Path Disclosure the BERT based models achieved better results than models based on the other two embeddings.

The rest of the paper is organized as follows.
In Section~\ref{sec:related} we introduce works that are similar to ours.
Section~\ref{sec:dataset} gives details about our dataset collection methodology, while in Section~\ref{sec:approach} we describe our overall approach for the systematic comparison of the different source code embedding methods.
Section~\ref{sec:results} contains the comparison results.
We summarize the set of threats to the validity of our work in Section~\ref{sec:threats} and conclude our findings in Section~\ref{sec:conclusion}.

\vspace{-10pt}
\section{Related Work}\label{sec:related}
\vspace{-5pt}

Solving a SE task with machine learning requires the input source code to be represented as a numeric vector.
Therefore many approaches have been proposed for deriving meaningful code representations to feed into ML models.

Alon et al.~\cite{alon2019code2vec} introduce code2vec, a neural model for representing snippets of code as continuous distributed vectors.
The method first breaks down the code to a collection of paths in its abstract syntax tree.
Then, the network learns the atomic representation of each path while simultaneously learning how to aggregate a set of them. 

Lozoya et al.~\cite{lozoya2021commit2vec} introduce a new code embedding technique called commit2vec based on code2vec.
This representation focuses on embedding code changes rather than code snapshots, which they used to successfully train models to detect vulnerability fixing commits.

Ben-Nun et al.~\cite{10.5555/3327144.3327276} propose a code embedding technique called inst2vec that is based on an Intermediate Representation (IR) of the code that is independent of the source programming language.
They provide a novel definition of \textit{contextual flow} for this IR, leveraging both the underlying data- and control-flow of the program.
The athors of the paper demonstrate the effectiveness of the approach on compute device mapping, optimal thread coarsening and algorithm classification .

Mou et al.~\cite{10.1007/978-3-319-25159-2_49} propose the ``coding criterion'' to build program vector representations, which are the premise of deep learning for program analysis.
They evaluate the learned vector representations both qualitatively and quantitatively.


In our work, we do not propose new code embedding techniques, rather evaluate the impact of different text representation methods (i.e., word embeddings) used as code representations for vulnerability prediction.
As these techniques are usually used under the hood of the more complex code representations, this is a natural first step towards better understanding the application of natural language processing techniques for solving SE tasks.

There are also many related works that focus on vulnerability prediction or similar SE tasks based on word embeddings in particular.
Harer et al.~\cite{harer2018automated} use word2vec to create word embeddings for C/C++ tokens.
Based on this code representation they successfully train machine learning models to predict the results of static analyzers.

White et al.~\cite{White_2019} apply word2vec in the scope of automatic program repair.
In their approach, DeepRepair, they create Java token embeddings that they use to start a recursive encoder for abstract syntax trees.
Chen and Monperrus~\cite{chen2018remarkable} use word2vec to create Java token embeddings for automated program repair as well, in order to find the correct ingredients.

Unlike the works mentioned above, we are not focusing on creating a state-of-the-art prediction model for a particular SE task.
Rather, we investigate the capabilities of various text vectorization techniques as source code representations in the context of identifying vulnerabilities in Python code with ML.

The work of Russel et al.~\cite{russell2018automated} is the closest one to ours.
They developed a fast and scalable vulnerability detection tool based on deep feature representation learning that directly interprets lexed source code.
They compared the bag-of-words (BOW) based simple source code embedding with code representations learned by CNN and RNN models automatically (i.e., with an embedding layer).
Although our approach is similar, we do not compare different ML models and automatic code representation learning, but explicitly compare the effect of applying word embeddings as features for an LSTM prediction model.
To the best of our knowledge, ours is the first attempt to systematically evaluate the impact of word2vec, fastText, and BERT-based code embeddings on vulnerability prediction.

\vspace{-10pt}
\section{Dataset Extraction}\label{sec:dataset}
\vspace{-6pt}

\begin{figure}
\vspace{-5pt}
\includegraphics[width=\textwidth]{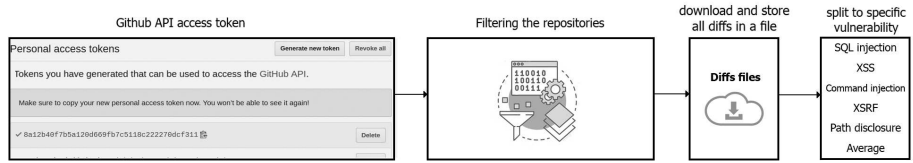}
\vspace{-20pt}
\caption{GitHub data mining process to find vulnerability fixes}\label{fig:fig2}
\vspace{-20pt}
\end{figure}

In order to compare the various word embedding based code representations, we need a training dataset for the vulnerability prediction model relying on them.
We applied a data mining approach to gather actual vulnerability fixes from various repositories and use them to train our model to recognize different patterns of security vulnerabilities in source code.
We chose GitHub as our data source as it contains a wide range of open source applications, including Python source code, which we focus on in this work.
We searched for commits in Python projects with some vulnerability related keywords included in their commit messages.
There are numerous types of security vulnerabilities in programming languages, but most of them are spread across languages.
We focus on some of the most popular vulnerability types, namely SQL and command injection, cross-site scripting, cross-site request forgery, remote code execution, and path disclosure.
Our overall data mining process is displayed in Figure~\ref{fig:fig2}.

\vspace{-4pt}
\subsection{Mining GitHub}
We followed the guidance provided by Chaturvedi et al.~\cite{chaturvedi2013tools}, who demonstrated how to use tools and datasets to mine database repositories and assist us in gathering data in this time-consuming task.
The first step is to collect a large number of commits that are candidate vulnerability fixes.
We searched for vulnerability fixing commits by querying GitHub data through its public REST API.
Because of GitHub constraints, we first had to extract a dataset containing commits coming from different language projects and then filter out data related to Python projects.
We ended up collecting approximately 70k commits yielding to 140k Python code snippets (vulnerable and fixed together) from 14k different Python projects.
To facilitate reproducibility, we published all the collected data and processing scripts in the form of an online appendix.\footnote{\url{https://doi.org/10.5281/zenodo.4703996}}

\vspace{-15pt}
\subsection{Filtering the Data}
\begin{figure}
\vspace{-17pt}
\includegraphics[width=\textwidth]{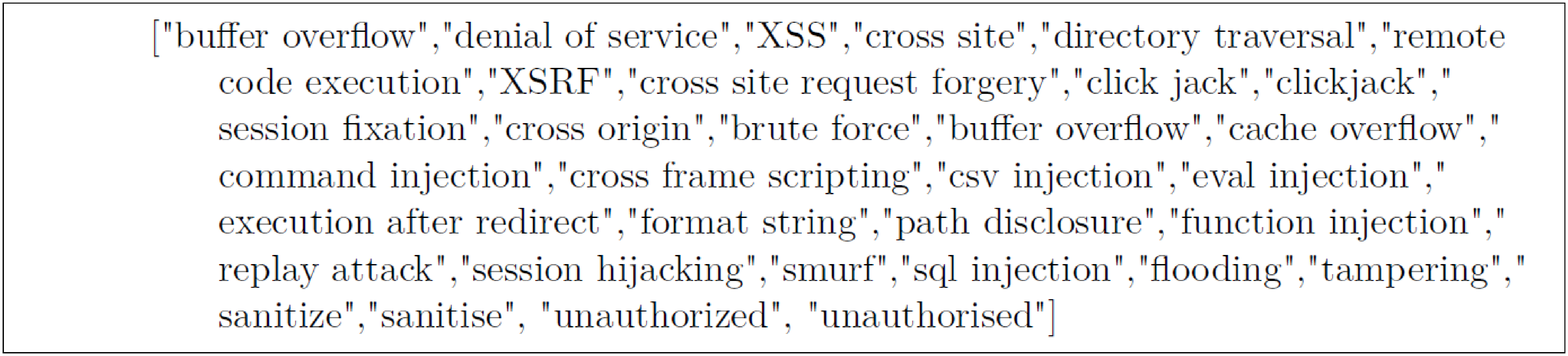}
\vspace{-22pt}
\caption{Security related search terms used by our heuristic data collection script} \label{fig:fig1}
\vspace{-20pt}
\end{figure}

\vspace{-7pt}
After collecting the candidate commits, we filtered them based on some security relevant keywords.
Some sample keywords we use in the heuristic data collection scripts are shown in Figure~\ref{fig:fig1}.
For the complete list of search terms, see the script source in the online appendix package.

We used the PyDriller tool~\cite{spadini2018pydriller} to download repositories and look for relevant commits and extract information from them.
We also filtered out commits that did not contain changes in files with '.py' extension.
Once we identified the commits that are related to vulnerability fixes, we downloaded the changed source code before and after the fixes.
It turned out that downloading the source code in a reasonable amount of time was possible if all of the scanning was done ahead of time in a clever way to keep the number of downloaded repositories to a bare minimum.
The diffs files we downloaded are essentially large text files that represent the changes in the source code introduced by a commit; however, they contained some unnecessary details (file name, line number) that we eliminated before assembling the final dataset using the previous and subsequent versions of the code snippet.
Both snippets contained the changed lines so we could extract and label the functions in the previous version as vulnerable while after the fix, they become not vulnerable.

\subsection{Labels}
After filtering the commits based on their messages and downloaded the relevant code changes in form of diffs, we created the final labeled dataset as follows.
We removed the comments from the affected code blocks because they are unlikely to impact a file's vulnerability.
After that, we extracted the fragments of code (i.e., code blocks) from the diff files that were affected by the fix in the commit and assigned the vulnerable label to its pre-fixed version while not vulnerable label was assigned to its fixed form.
However, it is not trivial to identify the exact code blocks within the whole source file that were affected by a vulnerability fixing change.
For this, we analyzed the downloaded diff files and implemented the algorithm presented by Hovsepyan et al.~\cite{hovsepyan2012software} and Wartschinski et al.~\cite{Wartschinski2014software} to find the appropriate code block.

\vspace{-10pt}
\section{Approach}\label{sec:approach}
\vspace{-5pt}

The primary goal of this work is to compare various embedding layers based on text representations in order to determine their capabilities in detecting vulnerabilities in Python programs.
To achieve an objective comparison, we need to apply the different source code embedding methods selected with the same ML algorithm trained on the same dataset.
We evaluate the embedding methods by training a Long Short-Term Memory (LSTM) model with the same hyper-parameters on the dataset described in Section~\ref{sec:dataset}.

\vspace{-10pt}
\subsection{The Evaluated Embedding Layers}
To encode the code tokens we need to transfer the code tokens into vectors using one of our selected embedding methods (word2vec, fastText and BERT).
For word2vec and fastText we need to train a model that learns the embeddings based on a large corpus of Python source code.
To collect this, we also mined GitHub for popular Python projects.

\vspace{-10pt}
\subsubsection{Word2vec:}

Word2vec is one of the most widely adopted word embedding methods to represent source code in vector form.
To derive word2vec based source code embeddings, we needed to train a suitable word2vec model on Python source code to encode the code tokens into word2vec vectors.
Training the word2vec model requires a large corpus of Python source code (for further reference, see the works of Bhoopchand et al.~\cite{bhoopchand2016learning} and Allamanis et al.~\cite{allamanis2013mining}).

\begin{figure}
\vspace{-20pt}
\includegraphics[width=0.9\textwidth]{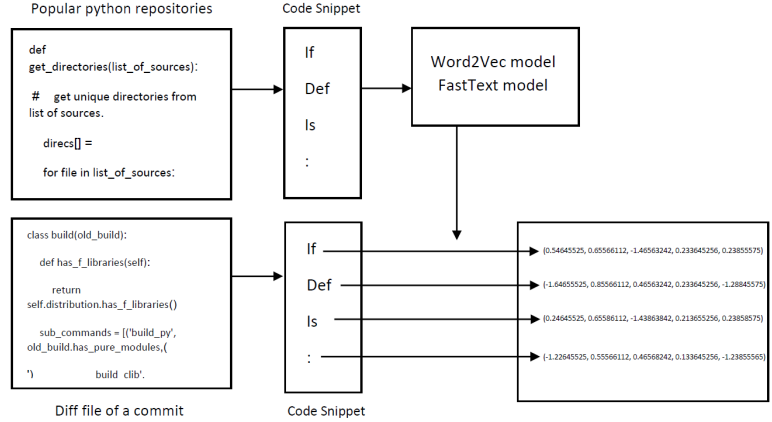}
\vspace{-5pt}
\caption{Transforming code into vectors} \label{fig:fig4}
\vspace{-40pt}
\end{figure}

To collect such a corpus, we searched for popular projects on GitHub.
GitHub uses two metrics to measure a repository's popularity: stars (user-created highlights) and forks (number of copies of a repository).
The list of selected repositories with a high number of stars and forks used as a code corpus is available in our online appendix.
We used PyDriller~\cite{spadini2018pydriller} for querying the most popular projects and downloading their source files.
The resulting source code, 11 million lines in total from 38 of the most popular projects, is simply concatenated to create a single massive Python code file.
Another script is then used to fix any issues with the text, such as indentation errors.
We transform the Python programs into Python tokens using the built-in Python tokenizer.
We delete the comments from files and add new lines at the end of the file.
Tabs and indentations have been normalized.
The word2vec model is then trained on the corpus using the Gensim~\footnote{\url{https://radimrehurek.com/gensim/}} Python package (see Figure~\ref{fig:fig4}).
All the word2vec training scripts are also part of our online appendix.

\subsubsection{FastText:}

We use the exact same process for calculating fastText~\cite{joulin2016fasttext} embeddings as for word2vec.
This means, we apply the same data analysis and tokenizer scripts and train the embedding model with the same Python code corpus.
We chose fastText as a study subject because word2vec only learns vectors for words that are complete in the training corpus.
FastText, on the other hand, learns vectors for both the n-grams and the full words contained inside each word.
FastText uses the mean of the target word vector and its component n-gram vectors for training at each step.
The change derived from the error is then applied uniformly to all of the vectors that were combined to form the target.
This adds a significant amount of extra computation to the training step.
A word must sum and average its n-gram component parts at each point.

\vspace{-15pt}
\subsubsection{BERT:}

Bidirectional Encoder Representations from Transformers (BERT) is a Google-developed Transformer-based machine learning technique for natural language processing (NLP) pre-training~\cite{devlin2018bert}.
Jacob Devlin and his Google colleagues developed and released BERT in 2018.
We selected this embedding method for comparison due to its recent successes within the NLP field.
As the BERT model is pre-trained on natural language texts, to adopt it to source code, we used its Microsoft's variant, called CodeBERT~\cite{feng2020codebert}.
CodeBERT is a pre-trained BERT model for programming languages.
In the context of this paper, we used BERT only as an embedding method and we feed all output tokens to an LSTM model.
The biggest difference between BERT and the other two embedding methods is that the training part of the embedding model is done in advance using a huge corpus in case of BERT, while for word2vec and fastText, we need to do the training locally.
This means that CodeBERT can be used out of the box, without having to train a model for token embeddings.

\subsection{Preparing the Data for Classification}

The collected vulnerability dataset (see Section~\ref{sec:dataset}) contains data in the form of vulnerable and not vulnerable code snippets.
We need to transform these into a list of tokens (such as '+', 'for', 'init') and convert each token into its vector representation according to the different embedding methods.
Each list of such vectors (representing the list of tokens of the underlying code snippet) will be labeled, where label '1' means that the code is vulnerable and '0' means it is not vulnerable.
Since ML models require a fix-sized input, we took the overall length of the concatenated vectors for the longest code snippet in our dataset and padded all the shorter ones with zeros to make them have the same length.

We split the training dataset into three sets, train, validation and test.
80\% of the data selected randomly is used for training, 10\% for validating and 10\% for testing.
Note that the validation data is only used to evaluate the model performance but all the results are presented on the test data, which the model has never seen before, applying a 10-fold cross-validation.

\vspace{-10pt}
\subsection{Training the LSTM}

After transforming each of our learning samples into a fix-sized numeric vector, we are ready for training the LSTM model on them.
For the implementation of the model, we used the Keras library~\cite{chollet2018keras}.
The first component of the model is the LSTM layer, which learns the features associated with the label of the code snippet (i.e., whether it is vulnerable or not).
We can use a variety of hyper-parameters for the model, like dropout.

Then, there is an activation layer that creates a dense output layer with one neuron.
We used the Sigmoid activation function as our aim is to predict between two classes: vulnerable and not vulnerable.
We applied different hyper-parameters and tried out several different combinations of them as a set.
Technically, the evaluation metric and loss functions are also hyper-parameters.
We chose to compare the model performances based on the F1-score metric.

\begin{figure}
\vspace{-20pt}
\includegraphics[width=\textwidth]{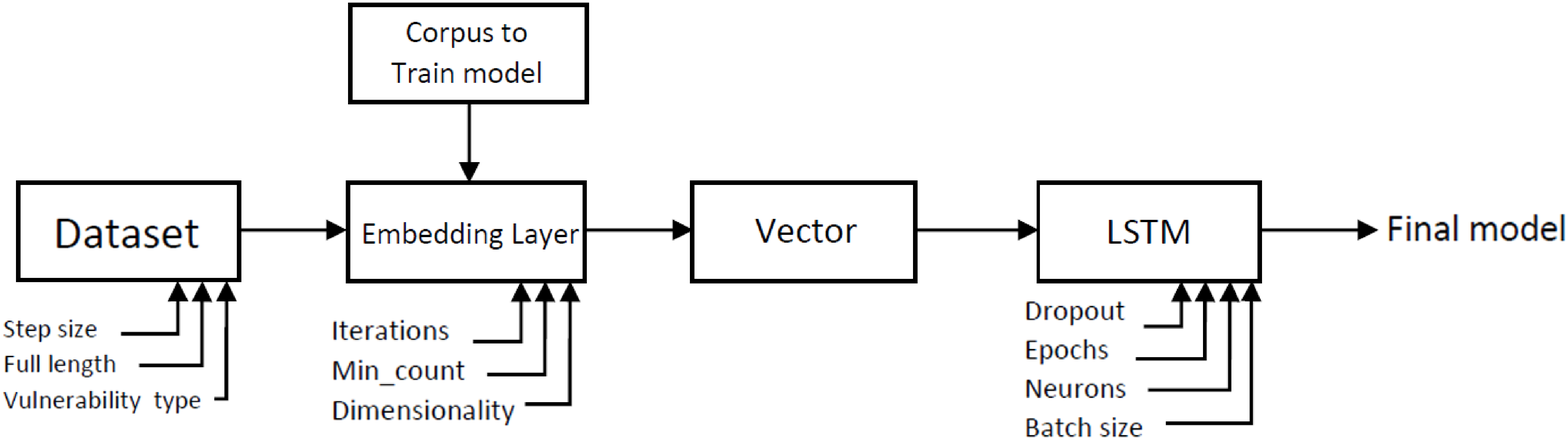}
\vspace{-20pt}
\caption{The steps of creating and evaluating the vulnerability prediction models based on different code representations} \label{fig:fig5}
\vspace{-15pt}
\end{figure}

Our model's base hyper-parameter is the number of neurons, which has a direct impact on its learning capacity; the more neurons we use, the more complex structures our model can recognize, but the training can also take longer.
Finally, we have the number of epochs, or the number of times the learning algorithm can iterate over the entire training data set, which we set to 100 and 200.
The high-level overview of our code representation evaluation/comparison process is shown in Figure~\ref{fig:fig5}.

\vspace{-10pt}
\section{Results}\label{sec:results}
\vspace{-5pt}

With data mining, we created a large dataset of Python code snippets from GitHub and labeled them as being vulnerable or not vulnerable based on detected vulnerability fixing commits.
The dataset covers six common types of vulnerabilities (SQL and command injection, cross-site scripting, cross-site request forgery, remote code execution, and path disclosure).
We trained an LSTM classifier using different embedding layers (i.e., word2vec, fastText, and BERT) with different hyper-parameters to compare the impact of the three different source code embeddings on vulnerability detection predicting performance.

\begin{figure}
\vspace{-20pt}
\includegraphics[width=\textwidth]{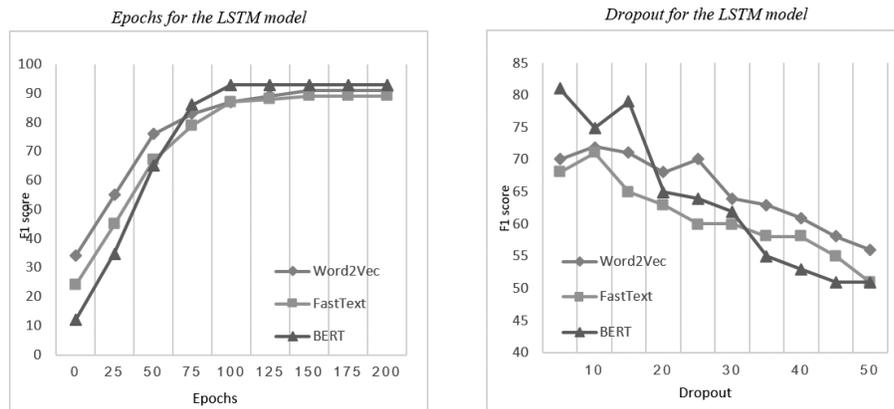}
\caption{LSTM model results using word2vec, fastText and BERT with various epochs and dropout ratio} \label{fig:lstm1}
\vspace{-10pt}
\end{figure}

\begin{figure}
\vspace{-5pt}
\includegraphics[width=\textwidth]{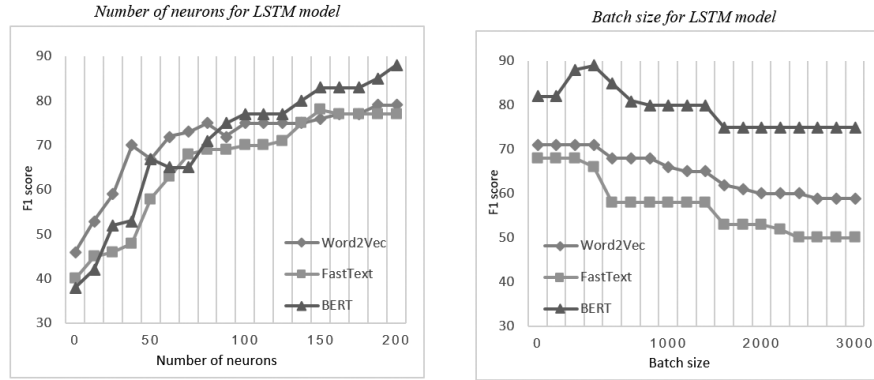}
\vspace{-14pt}
\caption{LSTM model results using word2vec, fastText and BERT with various number of neurons and batch sizes} \label{fig:lstm2}
\vspace{-14pt}
\end{figure}

The results of the LSTM models based on the three different code embedding layers are displayed in Figures~\ref{fig:lstm1} and \ref{fig:lstm2}.
Figure~\ref{fig:lstm1} shows the changes in F1 scores based on the number of epochs and ratio of dropout applied for the training.
As can be seen, the results are very close for the three embedding approaches.
Word2vec based results are slightly outperforming the others for small number of epochs and high droput rate.
However, for more than 75 epochs and a droput rate lower than 20\%, BERT based models perform the best.

Figure~\ref{fig:lstm2} depicts F1 score changes based on the hyper-parameters of neuron counts and batch sizes.
Again, the results achieved with the different embedding methods are very close.
For smaller number of neurons, word2vec based models work better, while for large number of neurons, BERT becomes superior to others.
BERT is also the best performing model when it comes to various batch sizes.
It works best with batch sizes between 400 and 600.
It is also true, that word2vec slightly outperforms fastText when applying larger batch sizes.



LSTM models using the BERT-based code representation achieve, on average, an accuracy of 93.8\%, a recall of 83.2\%, a precision of 91.4\%, and an an F1 score of 87.1\%.
The models based on the word2vec code representation achieve, on average, an accuracy of 91\%, a recall of 86.1\%, a precision of 88.2\%, and an F1 score of 85.6\%.
While fastText code representation based models achieve, on average, an accuracy of 91.8\%, a recall of 86.4\%, a precision of 85.1\%, and an F1 score of 84\%, on average, an accuracy of 91\%, a recall of 81\%, a precision of 90\% and an F1 score of 82\%.
Based on the results, we can answer RQ1 as follows.

\vspace{-8pt}
\begin{framed}
\vspace{-5pt}
\textbf{RQ1.} We did not observe significant differences in the vulnerability prediction performances of the LSTM models trained on different code embeddings. All of them are suitable the represent code for this task (all the models achieve an accuracy above 90\%). However, for BERT based models seems to perform slightly better, especially using larger batch sizes and smaller dropout rate.
\vspace{-5pt}
\end{framed}
\vspace{-8pt}

\begin{figure}
\vspace{-20pt}
\centering
\includegraphics[width=0.75\textwidth]{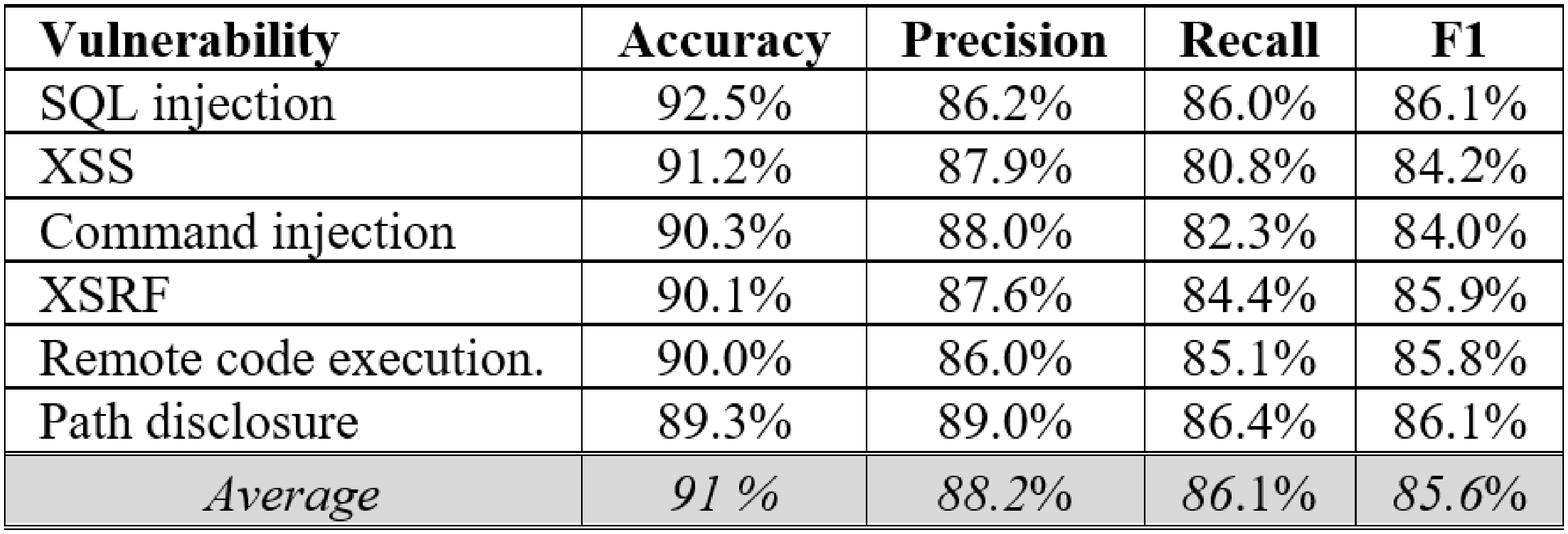}
\vspace{-10pt}
\caption{LSTM+word2vec results for each vulnerability categories}
\vspace{-20pt}
\label{fig:lstm-word2vec}
\end{figure}

\begin{figure}
\vspace{-20pt}
\centering
\includegraphics[width=0.75\textwidth]{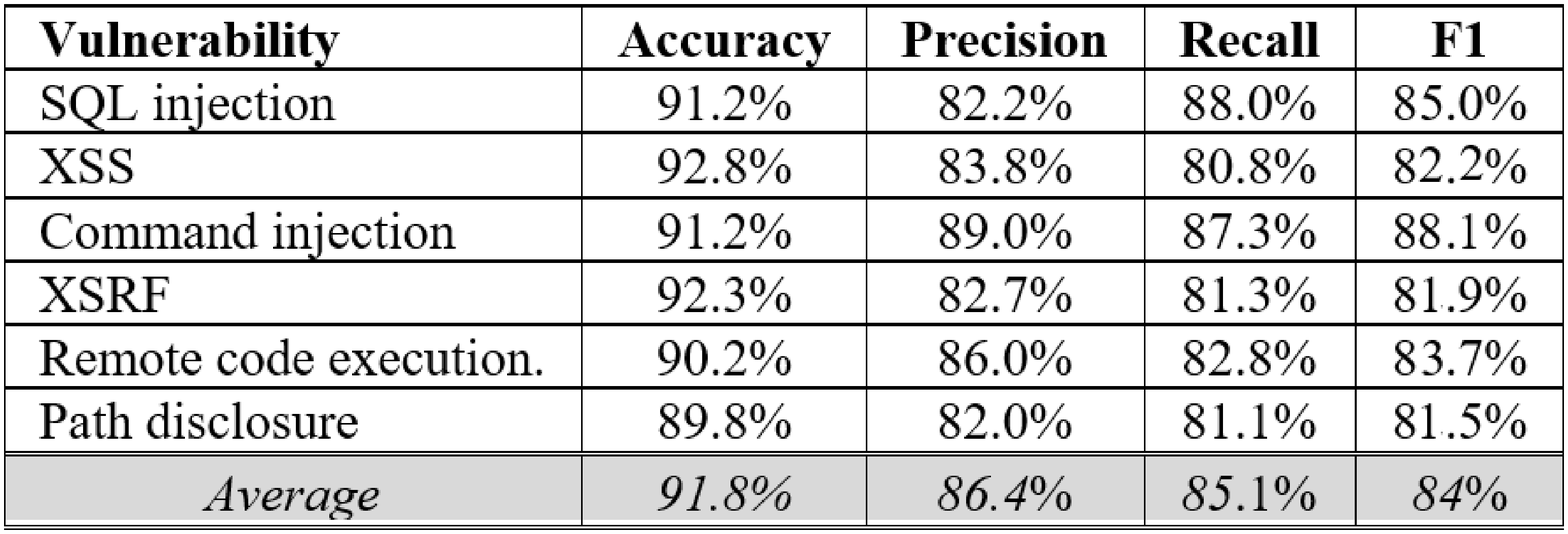}
\vspace{-10pt}
\caption{LSTM+fastText results for each vulnerability categories}
\vspace{-20pt}
\label{fig:lstm-fasttext}
\end{figure}

\begin{figure}
\vspace{-20pt}
\centering
\includegraphics[width=0.75\textwidth]{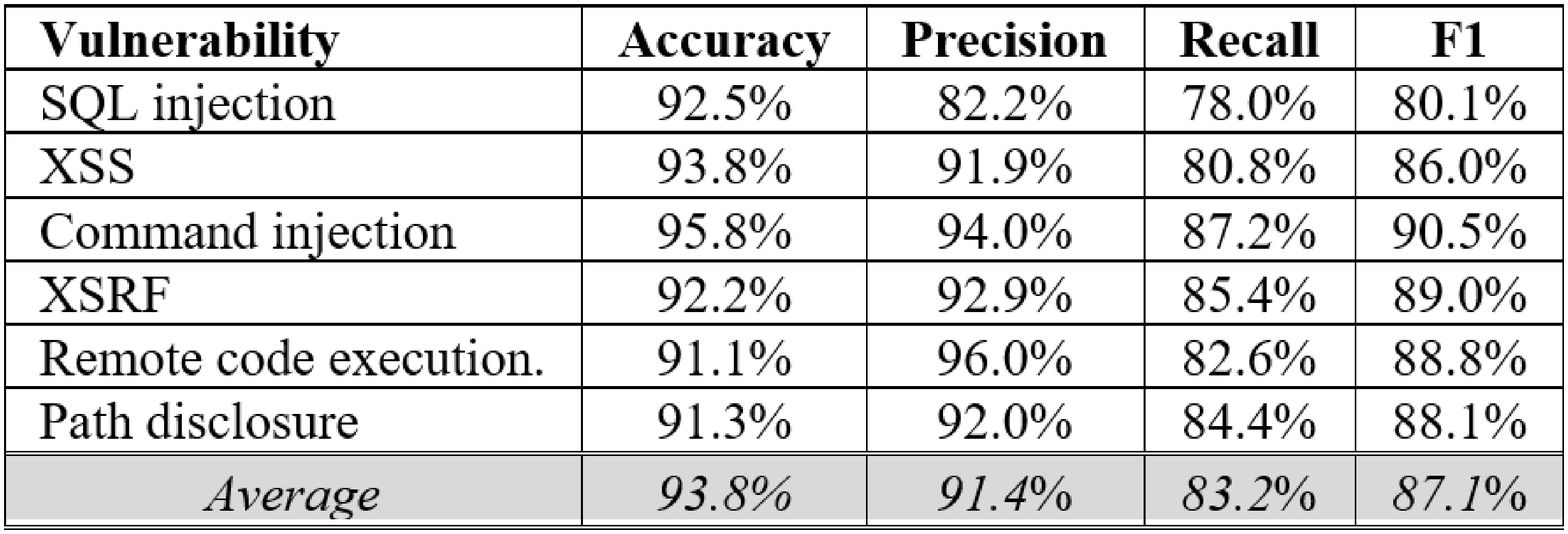}
\vspace{-10pt}
\caption{LSTM+BERT results for each vulnerability categories}
\vspace{-15pt}
\label{fig:lstm-bert}
\end{figure}

To answer RQ2, we calculated the same performance measures grouped by the different vulnerability types.
We categorized each code snippet according to the keywords found in the vulnerability fixing commit.
As it can be seen from Figures~\ref{fig:lstm-word2vec}, \ref{fig:lstm-fasttext}, and \ref{fig:lstm-bert}, there is not much variance in model performance values within the categories.

The word2vec based models (see Figure~\ref{fig:lstm-word2vec}) show the least variance in model performances within vulnerability categories.
The only minor exception is the recall for XSS, which is clearly lower than that of the others or the average, mostly because finding good vulnerable dataset of it is difficult and we think that we didn't train it with enough data.  
On the other hand, word2vec based models perform the best among all in detecting SQL injection vulnerabilities.

The fastText based models (see Figure~\ref{fig:lstm-fasttext}) show a higher variance within vulnerability categories.
They have similar average performance values to the word2vec based models, but are less effective in finding XSS, XSRF, and Path disclosure types of vulnerabilities.

The BERT based models (see Figure~\ref{fig:lstm-bert}) have the highest average performance measures and the variance in the values is small within the categories.
The only exception is SQL injection, where BERT based models are less efficient than the other models (recall of 78\%, F1 score of 80.1\%).
However, in case of Command Injection, Cross-Site Request Forgery (XSRF), Remote Code Execution (RCE), and Path Disclosure the BERT based models achieved better results than models based on other embeddings.
We can sum up the above observations to answer RQ2 as follows.

\vspace{-5pt}
\begin{framed}
\vspace{-5pt}
\textbf{RQ2.} The prediction models based on the different code representations show balanced performance measures within vulnerability categories. However, we found that vulnerability fix detection on top of word2vec-based embedded code outperform others in detecting SQL injection, while all the remaining vulnerability types are detected most effectively when BERT based models were used for code embedding	.
\vspace{-5pt}
\end{framed}
\vspace{-23pt}

\section{Threats to Validity}\label{sec:threats}
\vspace{-7pt}

The heuristic data collection is a major threat to the validity of the results.
With a keyword based commit search we might include irrelevant commits (that do not fix vulnerabilities) and we might miss out those that fix vulnerabilities but do not contain the searched keywords.
To mitigate this threat, we manually investigated a small sample of the collected data, which we found to be accurate in the majority of cases.
Since we collected a very large amount of such training data, the impact of several mis-classified commits should be negligible.

Many scenarios exist where a weakness arises from the interaction of lines of code that are distributed over a large file (or multiple files).
However, since the examples for vulnerabilities used to train the model only concentrate on the immediate vicinity of fixed lines, the model might be unable to learn the consequences of far-reaching dependencies.
Even though our results might not generalize to all of the different types of vulnerabilities, we believe these preliminary empirical results are already valuable. 

Limitations in the chosen approach may also be a threat to the internal validity.
We selected one specific prediction model to compare three different code embedding algorithms.
Different ML models might yield to different results.
In the future, we plan to extend our scope and add multiple ML models and code embeddings to the comparison.

\vspace{-13pt}
\section{Conclusion}\label{sec:conclusion}
\vspace{-7pt}


In this paper, we presented an empirical study where we performed the comparison of three word embedding based code representation methods in the context of vulnerability prediction in Python code.
These methods -- word2vec, fastText, and BERT -- adopted from the field of natural language processing, are widely used in practice to represent source code as numeric vectors and solve SE tasks (e.g., code summarization, bug detection, or finding copy-pasted code parts) with ML models trained on these representations.
Despite their popularity, very few works evaluate and compare their impact on the prediction performances of the ML models relying on them.

With a data mining approach, we collected vulnerability fixing commits from which we could extract vulnerable (before the fix) and not vulnerable (after the fix) code snippets that formed our training dataset (140k Python code snippets in total).
We applied the three investigated code embeddings to these code snippets and fed the resulting vectors into the same LSTM architecture to train a prediction model.
Our findings show that all the text representation methods are suitable for code representation in this particular task, but the BERT model is the most promising as it is the least time consuming and the LSTM model based on it achieved the best  overall accuracy(93.8\%) in predicting source code vulnerabilities.
Regarding the various vulnerability types, we observed slight variances in model performances based on the applied source code embeddings.
Nonetheless, the prediction model based on the word2vec representation of code clearly outperformed models based on fastText and BERT for detecting SQL injection, while in case of Command Injection, Cross-Site Request Forgery (XSRF), Remote Code Execution (RCE), and Path Disclosure the BERT based models achieved better results than models based on other embeddings.

Our future work will focus on using different classifiers and improving the approach for labelling the data, collecting a dataset of higher quality, and leveraging the commit context to create actionable fix recommendations.
The work could also be extended to other programming languages or types of vulnerabilities.

\vspace{-10pt}
\section*{Acknowledgment}
\vspace{-5pt}
The presented work was carried out within the SETIT Project (2018-1.2.1-NKP-2018-00004)\footnote{Project no. 2018-1.2.1-NKP-2018-00004 has been implemented with the support provided from the National Research, Development and Innovation Fund of Hungary, financed under the 2018-1.2.1-NKP funding scheme.} and supported by the Ministry of Innovation and Technology NRDI Office within the framework of the Artificial Intelligence National Laboratory Program (MILAB). The research was partly supported by the EU-funded project AssureMOSS (Grant no. 952647).

Furthermore, Péter Hegedűs was supported by the Bolyai János Scholarship of the Hungarian Academy of Sciences and the ÚNKP-20-5-SZTE-650 New National Excellence Program of the Ministry for Innovation and Technology.
\vspace{-10pt}

\bibliographystyle{splncs04}
\bibliography{references}

%
%
%
%
%
\end{document}